# Electrical switching of perpendicular magnetization in $L1_0$ FePt single layer

Liang Liu[1†], Jihang Yu[1†], Rafael González-Hernández[2,3], Jinyu Deng[1], Weinan Lin[1], Changjian Li[1], Chenghang Zhou[1], Tiejun Zhou[1], Herng Yau Yoong[1], Qing Qin[1], Han Wang[1], Xiufeng Han[4], Bertrand Dupé[2], Jairo Sinova[2*], Jingsheng Chen[1,5*]

[1]*Department of Materials Science and Engineering, National University of Singapore, Singapore 117575*

[2]*Institut für Physik, Johannes Gutenberg Universität Mainz, D-55099 Mainz, Germany*

[3]*Departament of Physics, Universidad del Norte, Barranquilla, Colombia*

[4]*Beijing National Laboratory for Condensed Matter Physics, Institute of Physics, Chinese Academy of Sciences, Beijing 100190, China*

[5]*Suzhou Research Institute, National University of Singapore, Suzhou, China, 215123*

[†]These authors contributed equally to this work.

*e-mail: msecj@nus.edu.sg; sinova@uni-mainz.de

**Electrical manipulation of magnetization is essential for integration of magnetic functionalities such as magnetic memories and magnetic logic devices into electronic circuits. The current induced spin-orbit torque (SOT) in heavy metal/ferromagnet (HM/FM) bilayers via the spin Hall effect in the HM and/or the Rashba effect at the interfaces provides an efficient way to switch the magnetization. In the meantime, current induced SOT has also been used to switch the in-plane magnetization in single layers such as ferromagnetic semiconductor (Ga,Mn)As and antiferromagnetic metal**

**CuMnAs with globally or locally broken inversion symmetry, respectively. Here we demonstrate the current induced *perpendicular* magnetization switching in *L*1$_0$ FePt single layer. The current induced spin-orbit effective fields in *L*1$_0$ FePt increase with the chemical ordering parameter (*S*). In 20 nm FePt films with high *S*, we observe a large charge-to-spin conversion efficiency and a switching current density as low as 7.0×10$^6$ A/cm$^2$. We anticipate our findings may stimulate the exploration of the spin-orbit torques in bulk perpendicular magnetic anisotropic materials and the application of high-efficient perpendicular magnetization switching in single FM layer.**

Electrical manipulation of magnetization through spin-orbit torque (SOT) has been widely investigated in multilayer heterostructures and bulk non-centrosymmetric conductors/semiconductors[1]. For multilayer heterostructures such as heavy metal/ferromagnet (HM/FM) bilayers, the SOT is considered to arise from the spin Hall effect (SHE)[2-4] in the HM[5-7] and/or the Rashba effect[8] at the interfaces[9-10]. According to the scenario of SHE, a spin current generated from HM is transferred to FM, exerting a spin transfer torque on FM. The torque efficiency is therefore significantly dependent on the spin current transmission transparency[11] and the spin relaxation[12] at the interface. The Rashba effect in HM/FM bilayers originates from the structural inversion asymmetry (SIA). With the Rashba spin-orbit interaction, non-equilibrium spin density of conduction electrons is generated in FM near the interface, which couples with the magnetic moments. For both effects in HM/FM bilayers, the switching current density ($J_c$) is proportional to the thickness ($t_{FM}$) of the FM layer due to the *interface* nature of the torque. Therefore, in order to achieve magnetization switching with low power consumption, an ultra-thin FM with a few atomic layer thickness is often used, which makes it difficult to achieve high thermal stability. For bulk non-centrosymmetric conductors/semiconductors, current induced magnetization switching has also been observed, such as in ferromagnetic semiconductor (Ga,Mn)As[13] (10-15 nm in thickness) with bulk inversion asymmetry (BIA) and in antiferromagnetic metals CuMnAs[14] (40-80 nm) and Mn$_2$Au[15] (75 nm) with locally broken inversion symmetry. For magnetization switching in these single layers, the current-induced non-equilibrium spin polarization generated uniformly in the *bulk* of the magnetic layer is able to exert a torque on local magnetic moments. Therefore, this switching behavior should not have thickness constraint. However, the current induced switching in single-layer (Ga,Mn)As, CuMnAs, and Mn$_2$Au is indeed magnetization rotation in the film plane, and the readout process utilizes the anisotropic magnetoresistance (AMR) effect, which is not conducive for ultra-high density storage.

In previous studies, it is commonly believed that the generation of SOTs would require either broken inversion symmetry[16] that leads to inverse spin galvanic effect (iSGE) or an extra layer (HM layer *et. al.*) with strong spin orbit coupling that leads to SHE. In this work, we demonstrate the observation of SOTs and current-induced *perpendicular* magnetization switching in $L1_0$ FePt single layer without extra heavy metal layer. We find the SOT in FePt has strong dependences on the thickness and the chemical ordering of the films, which cannot be simply interpreted by iSGE-SOT. In application, the high efficient electrical switching in $L1_0$ FePt will be of high interest to the industry because it enables memory cell with sufficient thermal stability to scale down to 3 nm due to its ultra-high perpendicular magnetic anisotropy[17].

6 nm and 20 nm $L1_0$-ordered FePt films with high perpendicular magnetic anisotropy (PMA) are epitaxially grown on SrTiO$_3$ (001) substrates. Details of sample preparation are presented in the Methods. The structural characterizations and magnetic properties of FePt films are given in Supplementary Section. Under the same deposition conditions, 6 nm FePt film has lower chemical ordering and therefore lower magnetic anisotropic field ($H_K$) than that of 20 nm FePt film. The films are patterned into 5 μm Hall bar for transport and optical measurements. Fig. 1 shows the current-induced magnetization switching in 6 nm $L1_0$ FePt

Hall bar. With a fixed in-plane magnetic field ($H_x$ = -1,000 Oe) along the negative current direction, we observe a deterministic magnetization switching behavior (Fig. 1a) by sweeping pulsed dc current. The polarity of the switching loop reverses after the magnetic field is reversed ($H_x$ = +1,000 Oe) as shown in Fig. 1b, indicating a typical SOT-induced switching behavior obtained in HM/FM bilayers. In order to more directly observe the magnetization switching process, we image the magnetization evolution of FePt under applying current by using a polar magneto-optic Kerr effect (MOKE) microscope. Same as in Fig. 1b, we apply a fixed positive in-plane magnetic field $H_x$ = 1,000 Oe. Figs. 1c-h and Figs. 1i-n show the magnetization evolutions after applying positive and negative pulsed current, respectively. Each image is taken 10 seconds after applying a current pulse to ensure that the magnetic morphology is in equilibrium state. The switching first happens at the center part of the current path (Fig. 1d and Fig. 1j), where the current density is higher than that in the Hall crossing area because of less shunting effect. The switched area gradually expands with increasing current density and finally fills the whole film (Figs. 1e-h and Figs. 1k-n). From the evolution, we conclude that the SOT induced switching in FePt films occurs via nucleation processes.

To systematically study the current-induced magnetization switching behavior in $L1_0$ FePt, we measure its field dependences in 6 and 20 nm films. Figs. 2a and 2b present the current-induced switching loops under various in-plane magnetic field $H_x$ ranging from -6,000 Oe to 6,000 Oe. We observe that the switching current density $J_c$ decreases with increasing $H_x$ as summarized in Fig. 2c, which is consistent with the standard SOT switching of perpendicular magnetization. However, it is surprising to see that $J_c$ for the 20 nm FePt film ($7.0\times10^6$ A/cm$^2$) is at least three times lower than that for the 6 nm FePt film ($2.6\times10^7$ A/cm$^2$). Whereas the 20 nm FePt film has a much larger magnetic anisotropic field than that of the 6 nm FePt film due to its higher chemical ordering. This indicates that the SOT in FePt is very sensitive to the thickness/chemical ordering. More interestingly, because of the multi-domain property in these micro-sized FePt Hall bar structures, intermediate magnetic states are observed. The stable multiple-level magnetic states induced by current in 6 nm FePt film is shown in Fig. 2d, which is consistent with the nucleation-type switching process as shown in Figs. 1f-h and Figs. 1l-n very and similar to that observed in antiferromagnetic materials[18,19] with multi-domain property.

Since the current-induced switching is usually accompanied with Joule heating, in order to more precisely estimate the charge-to-spin conversion efficiency in $L1_0$ FePt single layer, we conduct the Harmonic Hall measurement as well established for studying the SOT in HM/FM bilayers[20-23]. We apply a small ac excitation current and measure the in-phase first harmonic signal ($V_\omega$) and out-of-phase second harmonic signal ($V_{2\omega}$) simultaneously using two lock-in amplifiers. The current-induced Joule heating is very small (Supplementary Section). As shown in Fig. 3a, $\Delta H_L$ is defined to be the longitudinal component of the current induced effective magnetic field, while $\Delta H_T$ is the transverse component. The field dependences of $V_\omega$ and $V_{2\omega}$ are shown in Fig. 3c, e and Fig. 3d, f, respectively. The blue (orange) color in (Figs. 3c-f) corresponds to magnetization pointing along the +z-axis (-z-axis). The parabolic fits in Figs. 3c, e indicate coherent magnetization rotation during field sweeping. In Fig. 3d and Fig. 3f, non-zero values of $V_{2\omega}$ at zero field indicate contributions from anomalous Nernst effect (ANE) and other thermoelectric effects due to temperature gradient in the devices $\nabla T = (\nabla T_x, \nabla T_y, \nabla T_z)$[20,23,24]. After excluding these thermoelectric effects (Supplementary Section), we can estimate the SOT. With the planar Hall effect correction, the effective fields can be determined by the formula[22]: $\Delta H_{L(T)} = -2\dfrac{B_{L(T)} \pm 2\xi B_{T(L)}}{1-4\xi^2}$, where $B_{L(T)}$ is defined as $\left\{\dfrac{\partial V_{2\omega}}{\partial H} \Big/ \dfrac{\partial^2 V_\omega}{\partial H^2}\right\}_{H // L(T)}$. $\xi$ is the ratio of planar Hall resistance $\Delta R_{PHE}$ to anomalous Hall resistance $\Delta R_{AHE}$. The ± sign corresponds to magnetization pointing along the

±z-axis. The obtained current dependences of $\Delta H_L$ and $\Delta H_T$ for 20 nm FePt films are shown in Supplementary Section. The symmetries of $\Delta H_L$ and $\Delta H_T$ indicate a typical SOT property same as that studied in HM/FM bilayers. The amplitudes of $\Delta H_L$ and $\Delta H_T$ in 6 nm and 20 nm $L1_0$ FePt film under various current densities ($J$) are depicted in Fig. 3b. To compare the SOT with that in previous HM/FM bilayers, we calculate the spin torque efficiencies ($\beta_{L(T)}$), which is defined as $\Delta H_{L(T)}/J_e$. For the 20 nm FePt film, $\beta_L$ and $\beta_T$ are 60 Oe/ ($10^7$ A/cm$^2$) and 30 Oe/ ($10^7$ A/cm$^2$), respectively. While for the 6 nm FePt film, $\beta_L$ and $\beta_T$ are 12 Oe/ ($10^7$ A/cm$^2$) and 5 Oe/ ($10^7$ A/cm$^2$), respectively. The larger efficiencies $\beta_{L(T)}$ in the 20 nm FePt film is very likely to be the reason for its smaller switching current density in Fig. 2c. If we assume the longitudinal effective field $\Delta H_L$ comes from the current-induced spin current, the charge-to-spin conversion efficiency can be simply expressed as: $J_s/J_e = 2\dfrac{e\mu_0 M_s t_{FM}}{\hbar}\beta_L$. By using $M_s$ = 950 emu/cm$^3$ and $t_{FM}$ = 20 nm, we obtain $J_s/J_e$ = 3.46, which is one order of magnitude larger than that in HM/FM bilayers[20-23].

In $L1_0$ FePt, Fe and Pt atoms are alternatively stacked along the $c$ axis ([001] direction). In order to evaluate how this atomically layered crystal structure affects the SOT, 20 nm FePt films with different chemical orderings are fabricated under different temperatures. We observe that $\beta_L$ and $\beta_T$ increase with the chemical ordering parameter (S), as shown in Figs. 4a, b, which suggests that the SOT is closely related to the atomically layered structure ($L1_0$ phase). The 20 nm FePt film with the highest $S$ has the largest $\beta_{L(T)}$. In addition, we note that the 6 nm FePt film ($S$ = 0.49) has comparable $\beta_{L(T)}$ to that of the 20 nm FePt film with larger $S$ of 0.62, which implies that except for the chemical ordering an interfacial effect also contributes to the SOT in the 6 nm FePt film. It is worth noting that we also obtained comparable spin-orbit effective fields and current-induced switching (Supplementary Section) in $L1_0$ CoPt which is another member in the $L1_0$ family.

$L1_0$ ordered alloys (FePt and CoPt) have centrosymmetric bulk space group P4/*mmm* and

centrosymmetric polar site point group $D_{4h}$ for Fe (Co) and Pt atoms. With the symmetry consideration they should not have the SOC-induced non-equilibrium spin-polarization in the bulk form[16]. We carry out linear response Kubo-formalism calculations based on *ab-initio* calculations[25]. As for the thin films, the SOT of a finite FePt (001) slab with thickness of 2.2 nm (larger film thickness are not computationally possible) is obtained (Figs. 4c, d) and the details of the computation are given in Methods and Supplementary Section. In Figs. 4c, d, the calculated $β_L$ and $β_T$ are 3.1 Oe/ ($10^7$ A/cm$^2$) and 0.93 Oe/ ($10^7$ A/cm$^2$), respectively, which are close to our experimental results for the 6 nm FePt film. In addition, the calculated ratio $β_L/β_T$ is in good agreement with our experimental data. For the 20 nm FePt film, the calculation is much more difficult. For simplification, we calculate the SOT in the $L1_0$-FePt bulk phase and find it goes to zero. However, in our experiments, the 20 nm FePt film shows larger spin-orbit effective fields and smaller switching current density than that of the 6 nm film. This discrepancy between the experimental data and theoretical calculations implies that the the chemically ordered $L1_0$ FePt may have a novel type of SOT. New theory/mechanism is required to fully explain the experimental results.

Methods

*Sample growth and device fabrication.* 6 nm and 20 nm $L1_0$-odered FePt films are epitaxially grown on $SrTiO_3$ (001) substrates by dc magnetron sputtering (AJA) with Ar pressure of 10 mTorr and 50 W power. The sputtering rate of FePt target is 0.06 nm/s. After the deposition of FePt, the samples are left to cool to room temperature in the main sputtering chamber. 4 nm of $SiO_2$ film is then deposited as protection layer. VSM (Lake shore) is used to obtain the in-plane and out of plane magnetization hysteresis loops. All samples exhibit perpendicular magnetic anisotropy. After deposition, the films are coated with AZ1512 positive photoresist and patterned into 5 μm Hall bar structure by laser writer and Ar ion milling. Contact pads are defined by laser writer followed by the deposition of Ti (5 nm)/Cu (100 nm) and lift-off process. To create low-resistance electrical contacts between contact pads and the Hall bar, Ar ion milling is used to remove the $SiO_2$ protection layer just before deposition of Ti-Cu. The ready devices are then wire-bonded to sample holder for transport measurements (Quantum Design PPMS).

*Current-induced switching measurement.* For current-induced SOT switching measurement, a dc pulsed current $I_{pulse}$ with duration of 10 μs generated by Keithley 6221 is applied to the current path of the Hall bar. After each pulse (at least 4 seconds), a small ac excitation $I_{ac}$ (<10 μA) is output to the current path and the Hall voltage $V_{ac}$ is recorded by a lock-in amplifier. The Hall resistance characterizing the magnetization is defined as $R_H = V_{ac}/I_{ac}$.

*MOKE imaging measurements.* The magnetization of 6 nm $L1_0$ FePt film is imaged with a polar MOKE system. The white (black) color represents an overall up (down) magnetized state. An in-plane magnetic field (1,000 Oe) is applied along the Hall bar current path to break the rotational symmetry. The dc pulsed currents $I_{pulse}$ with duration of 10 μs are generated by Keithley 6221. Before applying positive pulses, a negative pulsed current with - $2.6×10^7$ A/cm$^2$ density is applied to initiate the magnetization direction. The MOKE image at this state is subtracted as a background. After subtraction, positive pulsed currents with density increasing from 0 to $2.6×10^7$ A/cm$^2$ are applied to show the magnetization switching scenario. The MOKE images are taken at least 10 seconds after the dc pulsed current is applied to ensure the moment is at a stable state. A reverse process is implemented for the switching experiment with negative pulses.

*Computational calculations. Ab-initio* calculations are performed based on the density functional theory (DFT) with generalized gradient approximation (GGA) in the form of Perdew-Burke-Ernzerhof (PBE)[27] functional as implemented in Vienna Ab-initio Simulation Package (VASP)[28]. We use an energy cutoff of 500 eV and Γ centered Monkhorst-Pack grids of 12×12×1 for k-points mesh. In all the calculations, the spin orbit coupling (SOC) is included self-consistently at the DFT level. In order to evaluate the magnitude of the effective field driving the spin-orbit torque, we calculate the current-induced spin polarization (CISP) using the Kubo linear response formalism[29] on an interpolated 1024×1024×1 k-grid. In this method, we follow the virtual crystal approximation to take into account finite disorder (conductivity) through a life-time broadening, Γ, that we match with the diagonal conductivity of the sample. In addition, an exchange coupling between the carriers and local moments is estimated to be 1 eV[29]. In the computation of the Kubo formula, we use the maximally localized Wannier function formalism[30] with 18 maximally localized Wannier functions (MLWFs) per atom. $L1_0$ FePt (001) films are simulated using a supercell with thirteen atomic layers separated by the vacuum gap of 11 Å. Four topmost layers of the slab are allowed to relax unconstrainedly until residual forces on all atoms reached 0.01 eV/Å.

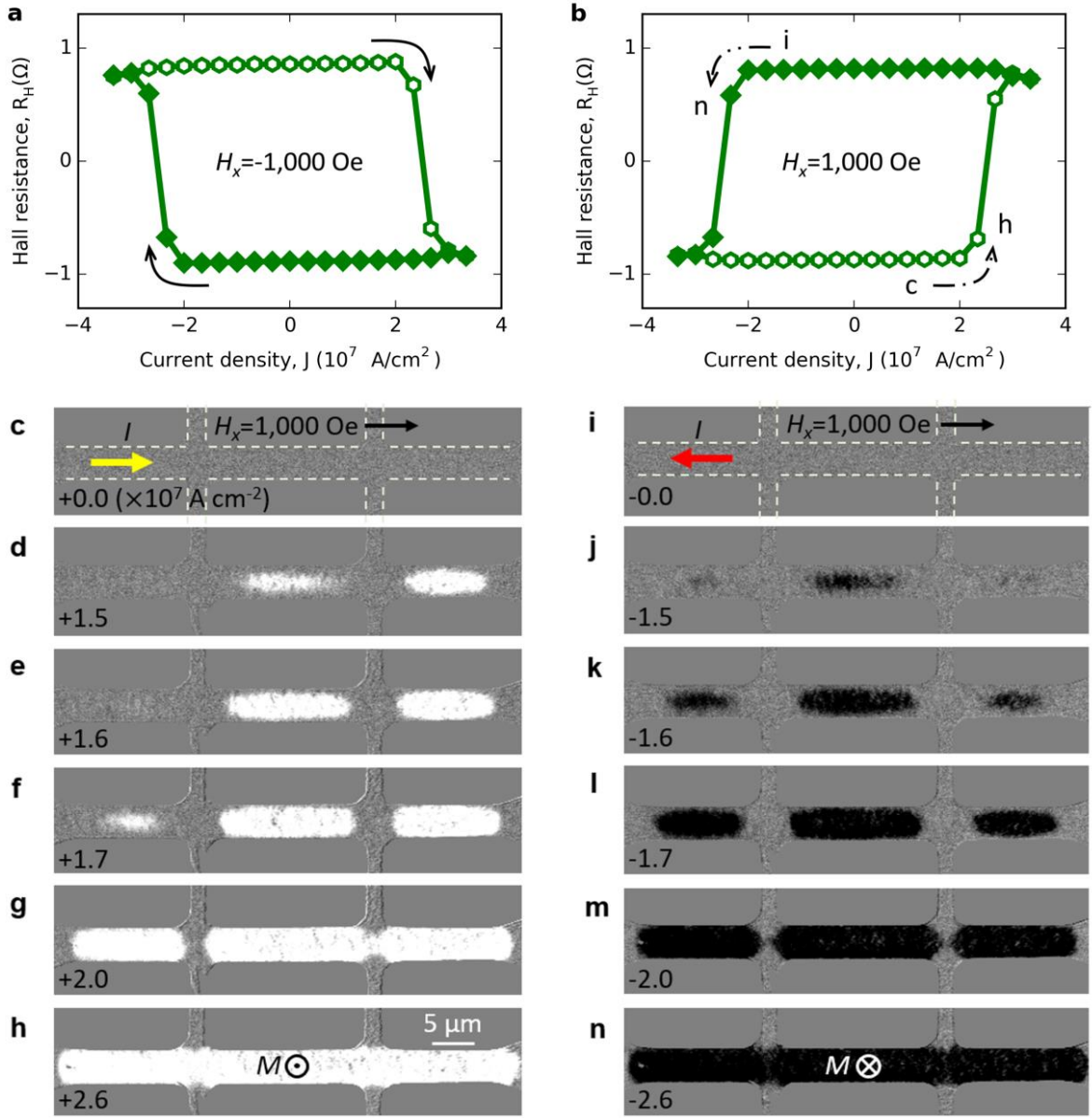

Fig. 1. Current-induced magnetization switching in $L1_0$ FePt single layer. a, b, Current-induced magnetization switching in 6 nm $L1_0$ FePt film in the presence of opposite in-plane magnetic fields, -1,000 Oe for (a) and 1,000 Oe for (b), respectively. c-h, Polar MOKE images showing the magnetization switching process by increasing the pulsed current amplitude from 0 to $2.6\times10^7$ A/cm$^2$. The light dashed lines in (c) indicate the Hall bar edges. A negative pulsed current with density of $-2.6\times10^7$ A/cm$^2$ is applied to initiate the magnetization before step (c). i-n, Polar MOKE images showing the magnetization switching

process by changing the pulsed current from 0 to -2.6×10$^7$ A/cm$^2$. A positive pulsed current with 2.6×10$^7$ A/cm$^2$ density is applied to initiate the magnetization before step (i). The pulse width for all sub-figures is 30 μs. The processes of (c-h) and (i-n) are indicated in the corresponding switching parts in (b).

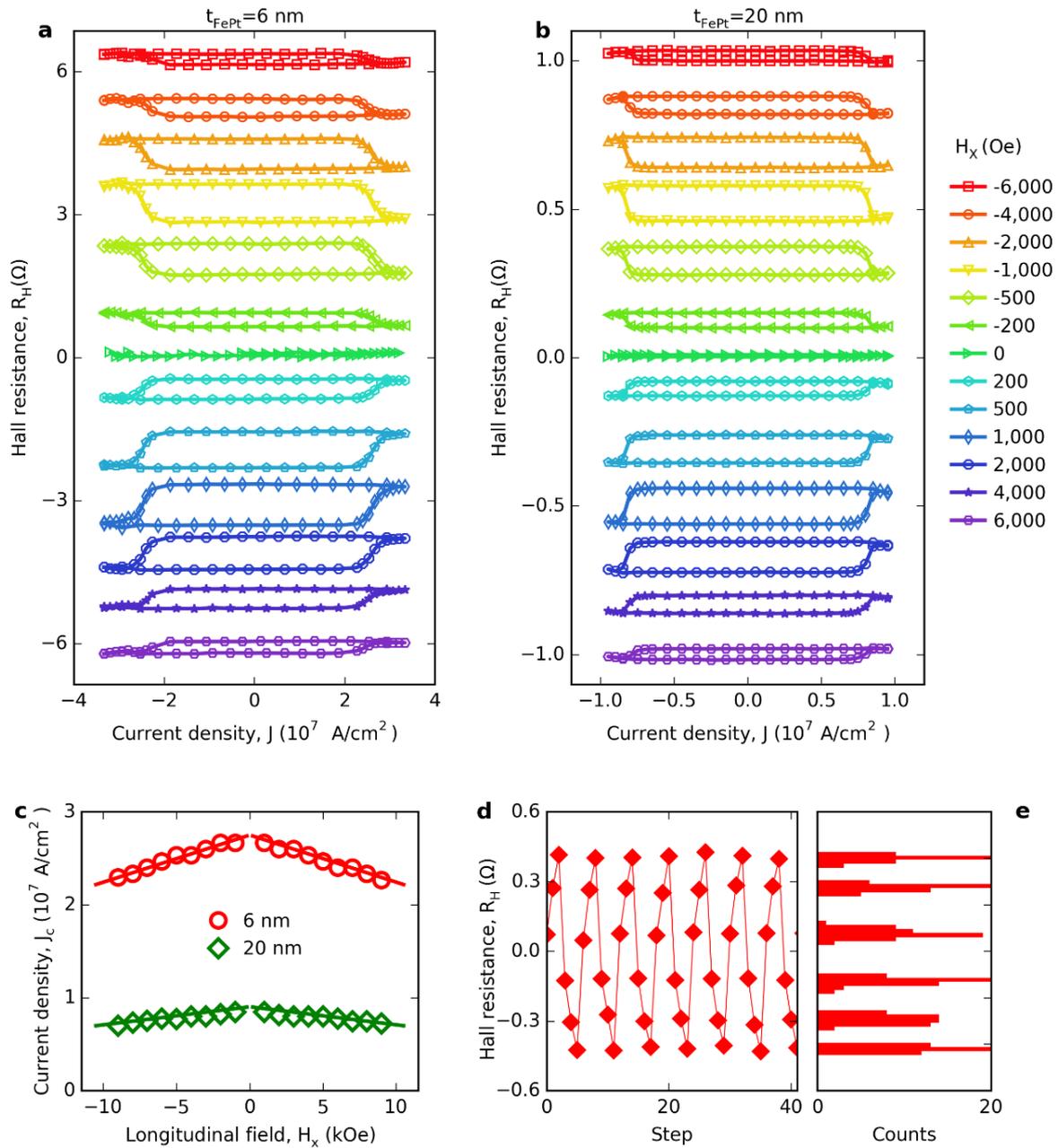

Fig. 2. Current-induced magnetization switching in 6 nm and 20 nm $L1_0$ FePt films under various in-plane magnetic fields. a, b, SOT switching loops obtained under various $H_x$ for 6 nm and 20 nm $L1_0$ FePt films, respectively. c, In-plane field dependence of critical switching

current density ($J_c$). The two solid lines are for eye guiding. d, Current-induced multi-level switching in 6 nm FePt film with controlled series of electrical current density. Six intermediate states (-0.45 Ω, -0.3 Ω, -0.15 Ω, 0.1 Ω, 0.3 Ω, 0.4 Ω) are evident. e, Histogram of the six resistance states in (d), obtained from 50 repetitions of "3 positive+3 negative" pulsed current sequence.

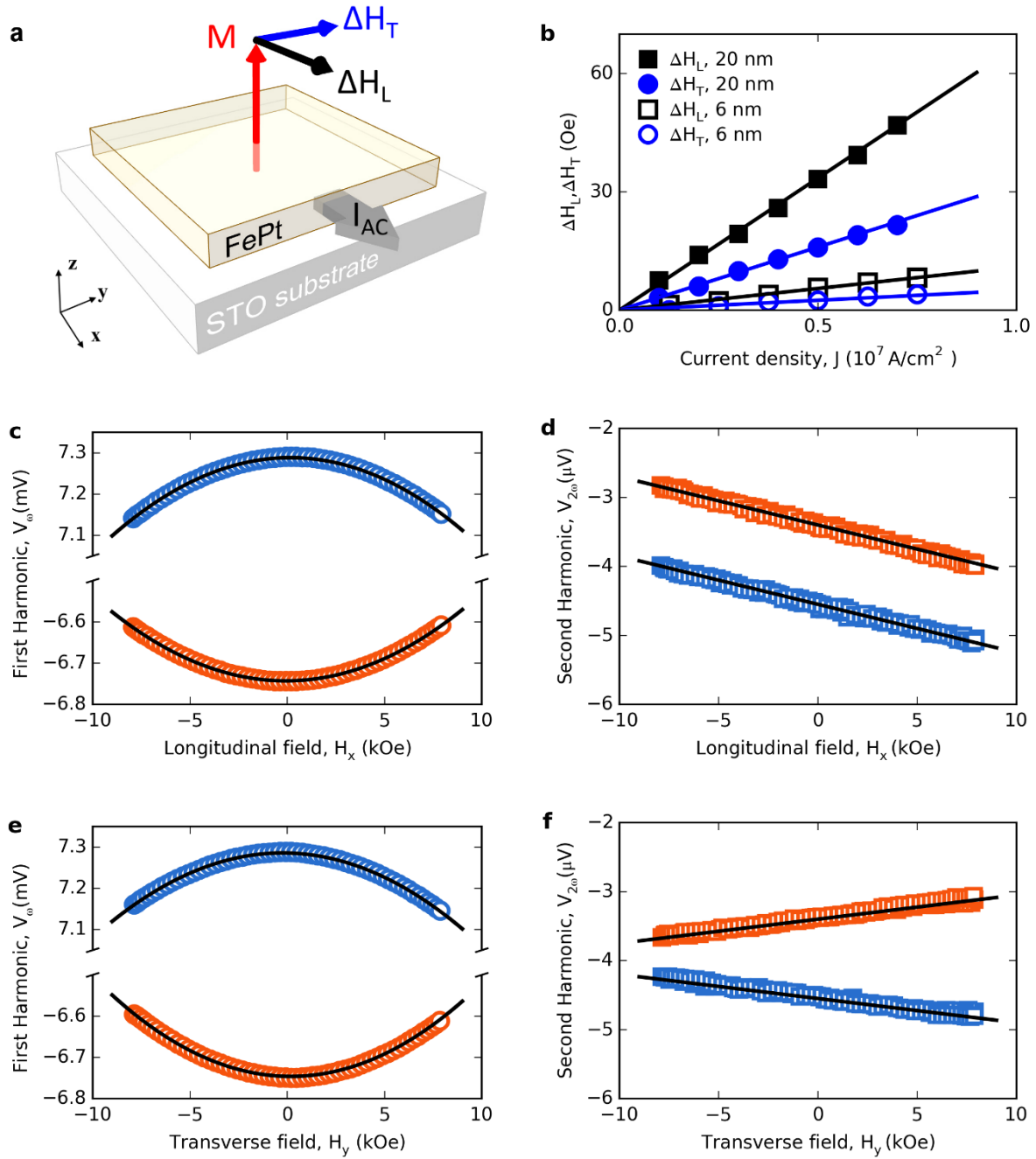

Fig. 3. Estimation of spin-orbit effective fields in $L1_0$ FePt films with harmonic Hall

measurements. a, Schematic illustration of the spin-orbit effective fields ($\Delta H_L$ and $\Delta H_T$) in $L1_0$ FePt films. b, Calculated effective fields at different current densities for 6 nm and 20 nm FePt films. Magnetic field dependence of first (c, e) and second (d, f) harmonic signals for 20 nm FePt film with 10 mA ac excitations. The external magnetic field $H_x$ and $H_y$ are swept along the longitudinal (c, d) and transverse (e, f) directions, respectively. The angle between the external field and the film plane is kept at 0 degree. The blue and orange colors in (c-f) corresponds to magnetization pointing along the +z-axis and -z-axis, respectively.

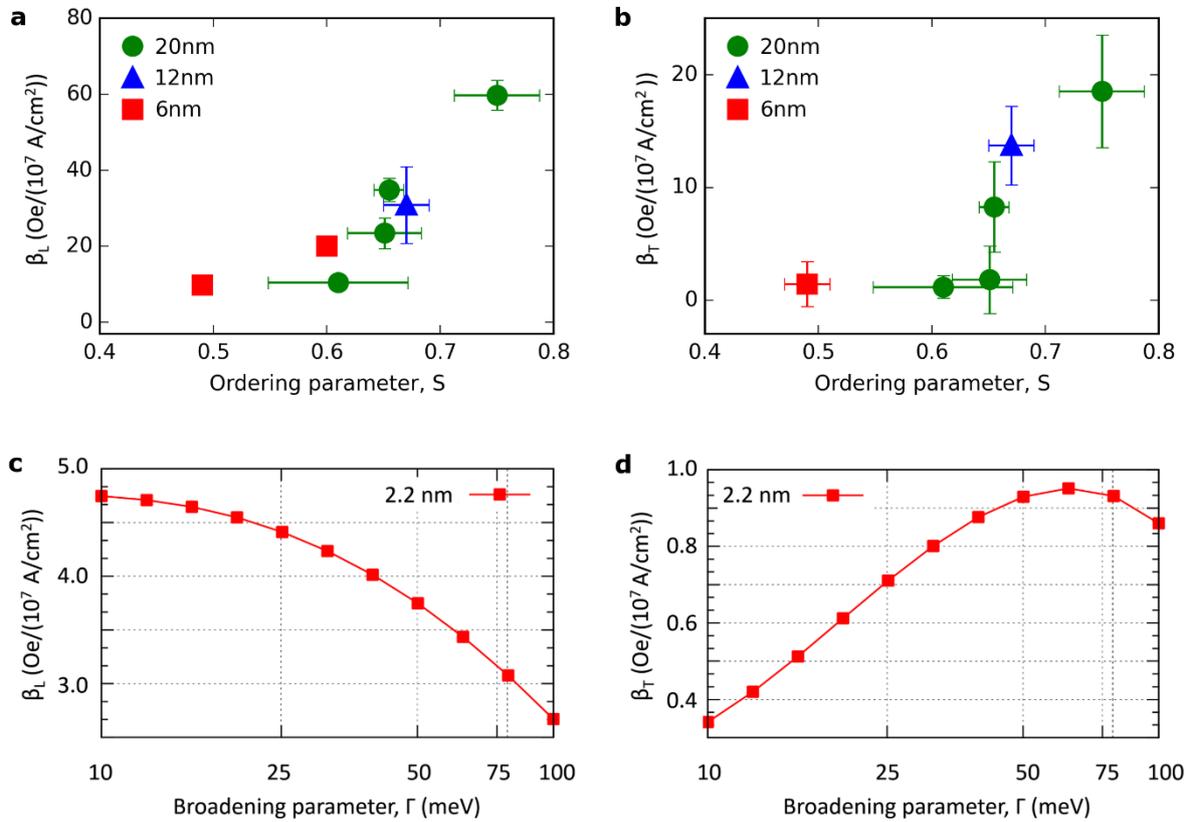

Fig. 4. Ordering dependence and theoretical calculations of spin-orbit effective fields in $L1_0$ FePt films. (a) Longitudinal and (b) transverse spin torque efficiencies of $L1_0$ FePt films with various chemical orderings (0.49 to 0.75) and thicknesses (6 nm, 12 nm, and 20 nm). The effective spin torque efficiencies for each points in (a, b) are obtained on at least 5 devices. The average value and standard deviation for the 5 devices are presented as data point and vertical error bar, respectively. The horizontal error bar comes from the errors in integration

of the peak areas during calculation of the ordering parameter. c, d, *Ab-initio* calculations of the longitudinal and transverse spin torque efficiencies of $L1_0$ FePt films versus life-time broadening parameter ($\Gamma$). $\Gamma$ = 79 meV is obtained by reproducing the electric conductivity.